\documentstyle[12pt]{article}

  \def\pp{{\mathchoice
          {%displaystyle
              \kern 1pt%
              \raise 1pt
              \vbox{\hrule width5pt height0.4pt depth0pt
                    \kern -2pt
                    \hbox{\kern 2.3pt
                          \vrule width0.4pt height6pt depth0pt
                          }
                    \kern -2pt
                    \hrule width5pt height0.4pt depth0pt}%
                    \kern 1pt
           }
            {%textstyle
              \kern 1pt%
              \raise 1pt
              \vbox{\hrule width4.3pt height0.4pt depth0pt
                    \kern -1.8pt
                    \hbox{\kern 1.95pt
                          \vrule width0.4pt height5.4pt depth0pt
                          }
                    \kern -1.8pt
                    \hrule width4.3pt height0.4pt depth0pt}%
                    \kern 1pt
            }
            {%scriptstyle
              \kern 0.5pt%
              \raise 1pt
              \vbox{\hrule width4.0pt height0.3pt depth0pt
                    \kern -1.9pt  %[e]=0.15pt
                    \hbox{\kern 1.85pt
                          \vrule width0.3pt height5.7pt depth0pt
                          }
                    \kern -1.9pt
                    \hrule width4.0pt height0.3pt depth0pt}%
                    \kern 0.5pt
            }
            {%scriptscriptstyle
              \kern 0.5pt%
              \raise 1pt
              \vbox{\hrule width3.6pt height0.3pt depth0pt
                    \kern -1.5pt
                    \hbox{\kern 1.65pt
                          \vrule width0.3pt height4.5pt depth0pt
                          }
                    \kern -1.5pt
                    \hrule width3.6pt height0.3pt depth0pt}%
                    \kern 0.5pt%}
            }
        }}

  \def\mm{{\mathchoice
                       {%displaystyle
                             \kern 1pt
               \raise 1pt    \vbox{\hrule width5pt height0.4pt depth0pt
                                  \kern 2pt
                                  \hrule width5pt height0.4pt depth0pt}
                             \kern 1pt}
                       {%textstyle
                            \kern 1pt
               \raise 1pt \vbox{\hrule width4.3pt height0.4pt depth0pt
                                  \kern 1.8pt
                                  \hrule width4.3pt height0.4pt depth0pt}
                             \kern 1pt}
                       {%scriptstyle
                            \kern 0.5pt
               \raise 1pt
                            \vbox{\hrule width4.0pt height0.3pt depth0pt
                                  \kern 1.9pt
                                  \hrule width4.0pt height0.3pt depth0pt}
                            \kern 1pt}
                       {%scriptscriptstyle
                           \kern 0.5pt
             \raise 1pt  \vbox{\hrule width3.6pt height0.3pt depth0pt
                                  \kern 1.5pt
                                  \hrule width3.6pt height0.3pt depth0pt}
                           \kern 0.5pt}
                       }}

\catcode`@=11
\def\un#1{\relax\ifmmode\@@underline#1\else
        $\@@underline{\hbox{#1}}$\relax\fi}
\catcode`@=12

\let\du=\du                     % dot-under
\def\a{\alpha}
\def\b{\beta}

\def\d{\delta}

\def\h{\eta}

\def\j{\psi}

\def\l{\lambda}
\def\m{\mu}
\def\n{\nu}

\def\p{\pi}
\def\q{\theta}
\def\r{\rho}
\def\s{\sigma}
\def\t{\tau}

\def\F{\Phi}

\def\L{\Lambda}

\def\ve{\varepsilon}

\def\cy{{\cal Y}}

\def\bo{{\raise-.5ex\hbox{\large$\Box$}}}               % D'Alembertian
\def\pa{\partial}                                       % curly d
\def\TH{{\raise.2ex\hbox{$\displaystyle \bigodot$}\mskip-4.7mu \llap H \;}}
\def\face{{\raise.2ex\hbox{$\displaystyle \bigodot$}\mskip-2.2mu \llap {$\ddot
        \smile$}}}                                      % happy face
\def\sp#1{{}^{#1}}                              % superscript (unaligned)
\def\leftrightarrowfill{$\mathsurround=0pt \mathord\leftarrow \mkern-6mu
        \cleaders\hbox{$\mkern-2mu \mathord- \mkern-2mu$}\hfill
        \mkern-6mu \mathord\rightarrow$}
\def\dvec#1{\vbox{\ialign{##\crcr
        \leftrightarrowfill\crcr\noalign{\kern-1pt\nointerlineskip}
        $\hfil\displaystyle{#1}\hfil$\crcr}}}           % <--> accent
\def\dt#1{{\buildrel {\hbox{\LARGE .}} \over {#1}}}     % dot-over for sp/sb
\def\frac#1#2{{\textstyle{#1\over\vphantom2\smash{\raise.20ex
        \hbox{$\scriptstyle{#2}$}}}}}                   % fraction
\def\sfrac#1#2{{\vphantom1\smash{\lower.5ex\hbox{\small$#1$}}\over
        \vphantom1\smash{\raise.4ex\hbox{\small$#2$}}}} % alternate fraction
\def\bfrac#1#2{{\vphantom1\smash{\lower.5ex\hbox{$#1$}}\over
        \vphantom1\smash{\raise.3ex\hbox{$#2$}}}}       % "
\def\afrac#1#2{{\vphantom1\smash{\lower.5ex\hbox{$#1$}}\over#2}}    % "
\def\[{\lfloor{\hskip 0.35pt}\!\!\!\lceil}
\def\]{\rfloor{\hskip 0.35pt}\!\!\!\rceil}
\def\Lag{{\cal L}}
\def\du#1#2{_{#1}{}^{#2}}
\def\ud#1#2{^{#1}{}_{#2}}

\def\fracm#1#2{\hbox{\large{${\frac{{#1}}{{#2}}}$}}}

\def\un{\underline}
\def\fracmm#1#2{{{#1}\over{#2}}}

\def\low#1{{\raise -3pt\hbox{${\hskip 0.75pt}\!_{#1}$}}}

\def\Dot#1{\buildrel{_{_{\hskip 0.01in}\bullet}}\over{#1}}
\def\dt#1{\Dot{#1}}

\newskip\humongous \humongous=0pt plus 1000pt minus 1000pt
\def\caja{\mathsurround=0pt}
\def\eqalign#1{\,\vcenter{\openup2\jot \caja
        \ialign{\strut \hfil$\displaystyle{##}$&$
        \displaystyle{{}##}$\hfil\crcr#1\crcr}}\,}
\newif\ifdtup

\def\ref#1{$\sp{#1)}$}

\parskip=\medskipamount                 % space between paragraphs (LaTeX)
\thicklines                         % thick straight lines for pictures (LaTeX)

\begin{document}

% =========================== title page ==========================

\thispagestyle{empty}               % no heading or foot on title page (LaTeX)

{\noindent UMDEPP 00--067 \hfill May 2000 }\\
{\noindent ITP--UH--08/00 \hfill hep-th/0005126
 }\\

\noindent
\vskip1.3cm
\begin{center}

{\Large\bf N=2 Super-Born-Infeld Theory Revisited~\footnote{Supported 
in part by NSF grant \# PHY--98--02551}}

\vglue.3in

Sergei V. Ketov \footnote{
On leave from: High Current Electronics Institute of the Russian Academy 
of Sciences, Siberian \newline ${~~~~~}$ Branch, Akademichesky~4, Tomsk 
634055, Russia}

{\it Department of Physics\\
     University of Maryland\\
     College Park, MD 20742, USA }\\
and \\
{\it Institut f\"ur Theoretische Physik\\
     Universit\"at Hannover, Appelstr.~2\\
     Hannover 30167, Germany}\\
\vglue.1in
{\sl ketov@itp.uni-hannover.de}
\end{center}
\vglue.2in
\begin{center}
{\Large\bf Abstract}
\end{center}

I discuss the symmetry structure of the N=2 supersymmetric extension of
the Born-Infeld action in four dimensions, and confirm its interpretation
as the Goldstone-Maxwell action associated with partial breaking of N=4 
extended supersymmetry down to N=2, by revealing a 
 hidden invariance of
the action with respect to two non-linearly realized 
supersymmetries and two spontaneously broken translations.
I also argue about the uniqueness
of supersymmetric extension
 of the Born-Infeld action, and its 
possible
relation to noncommutative geometry. 

\newpage

\section{Introduction}

In ref.~\cite{my} I proposed the N=2 supersymmetric extension of the 
four-dimensional {\it Born-Infeld} (BI) action. I interpreted it as
the Goldstone-Maxwell action associated with  spontaneous (partial) breaking
of (rigid) N=4 supersymmetry down to N=2, and the N=2 (abelian) vector 
supermultiplet of Goldstone fields. The basic idea behind this interpretation
was the anticipated equivalence (modulo a non-linear field redefinition)
between the N=2 super-BI action in four dimensions and the gauge-fixed 
world-volume action of a D3-brane propagating in six dimensions. This 
equivalence was verified in ref.~\cite{my}, in the leading and subleading
orders only (see ref.~\cite{tr} too), while no direct argument was presented.
In this Letter I report on a progress in obtaining 
the transformation laws of the hidden non-linearly
realized symmetries (including spontaneously broken translations and extra
N=2 supersymmetry) which 
determine the form of the N=2 super-BI
action and prove its Goldstone nature. The 
uniqueness of 
N=2 superextension of the BI action is also discussed. I give an N=2
superconformal extension of the BI theory, and speculate about its possible
relation to noncommutative geometry. 
 
\section{Featuring the bosonic BI action} 

In this introductory section I recall some well-known facts about the bosonic
BI action, in order to provide a basis for the subsequent discussion of the
N=2 supersymmetric extension in sect.~3.

The bosonic BI action in flat four-dimensional spacetime with Minkowski 
metric $\h_{\m\n}$, $\m,\n=0,1,2,3$, reads~\footnote{The overall normalization
of the BI action yields the Maxwell term, $-\fracm{1}{4}F_{\m\n}F^{\m\n}$, 
as the leading \newline ${~~~~~}$ contribution. The D3-brane action has, in 
addition, the 
inverse string coupling constant in front \newline ${~~~~~}$ of the action.}
$$ S_{\rm BI}=-\fracmm{1}{b^2}\int d^4x\,\sqrt{-\det\left(
\h_{\m\n}+bF_{\m\n}\right)}~~,\eqno(1)$$
where $F_{\m\n}=\pa_{\m}A_{\n}-\pa_{\n}A_{\m}$, and $b>0$ is the dimensionful
parameter. For instance, in string theory one has $b=2\p\a'$, whereas in 
N=1 supersymmetric QED one has $b=e^2/(2\sqrt{6}\p m^2)$. In what follows, I
choose $b=1$ for simplicity.

The BI theory (1) can be thought of as the particular covariant deformation of
Maxwell electrodynamics by higher order terms depending upon $F$ only. In fact,
the BI theory also shares with the Maxwell theory some other physical 
properties, such as causal propagation, positive energy density and 
electric-magnetic duality (see, e.g., refs.~\cite{my,tr} and references 
therein). Unlike the Maxwell theory, its BI generalization gives rise to the
celebrated taming of the Coulomb self-energy , i.e. it smears the singularity
associated with a point-like charge in classical electrodynamics. Supersymmetry
is known to be compatible with causality, positive energy and duality, so that
one expects from supersymmetric BI actions the similar (properly generalized)
properties. It is indeed the case for the N=1 BI action \cite{n1}, and it 
should be the case for the N=2 BI action \cite{my} too. 

As is also quite clear from its origin, either in open string theory or in N=1 
scalar QED, the BI action is the {\it effective} action obtained by summing up
certain quantum corrections (to all orders in $b$) that are independent upon
spacetime derivatives $(\pa F)$ of the Maxwell field strength $F$. The 
effective action is dictated by S-matrix, being defined modulo local field
redefinitions. This does not, however, make the BI action to be ambiguous
since it depends upon the vector gauge potential $A$ only via its field 
strength $F$, while any local reparametrization of $A$ merely results in the
additive $\pa F$-dependent terms which
 are to be disregarded by
definition of the BI action,
$$ \d S=\int d^4x\,\d \Lag(F)=\int d^4x\,\fracmm{\pa\Lag}{\pa F_{\m\n}}
\pa_{\m}\d A_{\n}=-\int d^4x\,\pa_{\m}\fracmm{\pa\Lag}{\pa F_{\m\n}}\d A_{\n}
=O(\pa F)~.\eqno(2)$$
In other words, the BI action is the effective action of slowly varying (but
not necessarily small) abelian gauge fields, which is dependent upon $F$, being
independent upon $\pa F$. In supersymmetric BI theories the r\^ole of $F$ is
played by the gauge superfield strength $W$, so that the super-BI actions in
superspace are defined modulo spacetime derivatives of $W$.

\section{N=2 BI action and its symmetries}

The N=1 BI action is well-known to be the Goldstone-Maxwell action 
associated with spontaneous partial supersymmetry breaking N=2$\,\to\,$N=1 
and the N=1 vector supermultiplet of Goldstone fields
\cite{n1,tr}. Both N=1 and N=2 gauge
field theories are most  naturally  formulated in superspace, with manifest
off-shell N=1 or N=2 supersymmetry, respectively,  which makes a study of
partial breaking N=2$\,\to\,$N=1
rather straightforward, by starting from a
linear off-shell realization of N=2 supersymmetry and imposing a non-linear 
constraint. Partial breaking N=4$\,\to\,$N=2 
in N=2 superspace is more complicated
since a natural (off-shell and N=4 supersymmetric) formulation of N=4 gauge
theories does not exist.

The N=2 supersymmetric BI action can be formulated in the standard N=2 
superspace parametrized
by $Z=(x^{\a\dt{\a}},\q^{\a}_i,\bar{\q}^i_{\dt{\a}})$, where $\a=1,2$ and
$i=1,2$. The N=2 flat superspace covariant derivatives $(\pa_{\a\dt{\a}},
D^i_{\a},\bar{D}^{\dt{\a}}_i)$ satisfy the algebra
$$ \{ D^i_{\a},\bar{D}_{\dt{\a}j}\}=-2i\d^i_j\pa_{\a\dt{\a}}~,\quad
\{ D^i_{\a},D^j_{\b}\}=\{\bar{D}^{\dt{\a}}_i,\bar{D}^{\dt{\b}}_j\}=0~.
\eqno(3)$$
The standard realization is given by
$$ D^i_{\a}=\fracmm{\pa}{\pa \q^{\a}_i}+i\bar{\q}^{\dt{\a}i}\pa_{\a\dt{\a}}~,
\quad \bar{D}_{\dt{\a}i}=-\fracmm{\pa}{\pa\bar{\q}^{\dt{\a}i}}-i\q^{\a}_i
\pa_{\a\dt{\a}}~~.\eqno(4)$$
 
The $SL(2,{\bf C})$ and $SU(2)$ indices are raised and lowered by the use of 
Levi-Civita ($\ve$) symbols, as usual. I use the notation \cite{book}
$$ D^{ij}= \fracm{1}{2}D^{\a i}D^j_{\a}~,\quad D_{\a\b}=\fracm{1}{2}D_{\a i}
D^i_{\b}~,\quad (D^3)^{\a}_i=\fracmm{\pa}{\pa D^i_{\a}}D^4~,\quad
D^4=\prod_{\a,i}D^i_{\a}~,\eqno(5)$$
and similarly for $\bar{D}$'s and $\q$'s ($\bar{\q}$'s). The standard 
(Berezin)  integration rules imply
$$ \int d^4x d^8\q\,\Lag\equiv \int d^4x d^4\q d^4\bar{\q}\,\Lag=
\int d^4xd^4\q\,\bar{D}^4\Lag=\int d^4xd^4\bar{\q}D^4\Lag~.\eqno(6)$$

The abelian N=2 superfield strength is described by an N=2 restricted chiral
superfield $W$ satisfying  the off-shell N=2 superspace constraints
$$ \bar{D}^{\dt{\a}}_iW=0\quad {\rm and}\quad D^4W=\bo \bar{W}~.\eqno(7)$$
The second constraint (7) is just the N=2 Bianchi identity that implies
$\bo(D_{ij}W-\bar{D}_{ij}\bar{W})=0$ and, hence, 
$D_{ij}W=\bar{D}_{ij}\bar{W}$.~\footnote{The fields are assumed to vanish at
infinity.} A solution to  eq.~(7) in components reads (in N=2 chiral 
superspace parametrized by $y^{\m}=x^{\m}-\fracm{i}{2}\q^{\a}_i
\s^{\m}_{\a\dt{\a}}\bar{\q}^{\dt{\a}i}$ and $\q^{\a}_i$)
$$\eqalign{
 W(y,\q) &~=
 a(y) +\q^{\a}_i\j^i_{\a}(y)-\fracm{1}{2}\q^{\a}_i
(\vec{\t})\ud{i}{j}
\q^j_{\a}\cdot\vec{D}(y) \cr
&~ -i(\q^3)^{i\a}\pa_{\a\dt{\b}}\bar{\j}^{\dt{\b}}_i(y)+\q^4\bo\bar{a}(y)~,\cr}
\eqno(8)$$
where I have introduced the complex (physical) scalar $a$, the chiral 
(physical) spinor isodoublet $\j^i_{\a}$, the real (auxiliary) isotriplet
$\vec{D}=\fracm{1}{2}(\vec{\t})\du{i}{j}D\ud{j}{i}$, and the Maxwell field
strength $F_{\m\n}$ subject to the Bianchi identity
$$ \ve^{\m\n\l\r}\pa_{\n}F_{\l\r}=0~,\eqno(9)$$
whose solution is just $F_{\m\n}=\pa_{\m}A_{\n}-\pa_{\n}A_{\m}$.  

The N=2 supersymmetric extension of the BI action, proposed in ref.~\cite{my}, 
reads
$$ S= \fracm{1}{2} \int d^4x d^4\q\, W^2 +\fracm{1}{8}\int d^4x d^8\q\,
 \cy(K,\bar{K})W^2\bar{W}^2~,\eqno(10)$$
where $K=D^4W^2$ and $\bar{K}=\bar{D}^4\bar{W}^2$, and
$$\eqalign{
 \cy(K,\bar{K})&~=~ \fracmm{1-\frac{1}{4}(K+\bar{K})-\sqrt{(1-\frac{1}{4}K
-\frac{1}{4}\bar{K})^2-\frac{1}{4}K\bar{K}}}{K\bar{K}}\cr\
&~=~ 1+\frac{1}{4}(K+\bar{K})+O(K^2)~.\cr}
\eqno(11)$$
The action (10) can be rewritten to the form
$$ S=\frac{1}{4}\int d^4x d^4\q \, X +
\frac{1}{4}\int d^4 x d^4\bar{\q}\,\bar{X} +O(\pa W)~,\eqno(12)$$
where the N=2 chiral lagrangian $X$ is the iterative solution to the N=2 
non-linear constraint \cite{my}
$$ X = \fracm{1}{4}X\bar{D}^4\bar{X} +W^2~~.\eqno(13)$$

The uniqueness of the N=2 BI action was questioned in ref.~\cite{kuz} by
presenting a calculation of some terms in the iterative solution to eq.~(13), 
which are absent in the perturbative expansion of the action (10), for example,
$$ \int d^4x d^8\q\,W^2\bar{W}^2\left[ (D^4W^2)\bar{D}^4(\bar{W}^2D^4W^2)
+(\bar{D}^4\bar{W}^2)D^4(W^2\bar{D}^4\bar{W}^2)\right]\eqno(14a)$$
versus
$$ \int d^4x d^8\q\,W^2\bar{W}^2\left[ (D^4W^2)^2\bar{D}^4\bar{W}^2
+(\bar{D}^4\bar{W}^2)^2D^4W^2\right]~.\eqno(14b)$$
However, it is not difficult to verify, by the use of eqs.~(3) and (7), 
that the difference between eqs.~(14a) and (14b) amounts
to the $\pa W$-dependent terms which do not belong to the N=2 BI action
 because they are ambiguous ({\it cf.} sect.~2). It was also explicitly
demonstrated in ref.~\cite{kuz} that the N=2 BI action (12) is self-dual with
respect to an N=2 supersymmetric electric-magnetic duality (claimed in 
ref.~\cite{my} too), by keeping all terms in the solution to eq.~(13), 
including the $\pa W$-dependent ones. This means that taking into 
account {\it some} $\pa W$-dependent terms is apparently needed to demonstrate
the N=2 supersymmetric electric-magnetic  duality of the N=2 BI action. 
In general,
however, it does not make sense to keep {\it some} $\pa W$ (or $\pa F$)
dependent  terms in the effective BI action originating either from a quantized
open superstring theory or from a quantized supersymmetric gauge theory, while
ignoring {\it other} possible $\pa W$ (or $\pa F$) dependent quantum 
corrections. Perhaps, the N=2 electric-magnetic self-duality may, 
nevertheless, be useful for a study of derivative corrections to the N=2 BI 
action in a more fundamental framework than just N=2 supersymmetry.

The Goldstone interpretation of the N=2 BI action implies that the complex
scalar $\left.W\right|=a=P+iQ$  is the Goldstone field associated with  two
spontaneously broken translations (in the directions orthogonal to a D3-brane
world-volume in six dimensions).  Hence, the action (10) or (12) should possess
hidden invariance  with respect to spontaneously broken (non-linearly
realized) translations, $\d a=\l +\ldots,$ where $\l$ is the complex (rigid)
parameter. This symmetry is obvious from the viewpoint of a (1,0) 
supersymmetric BI action in six dimensions \cite{my}, which is related to the
four-dimensdional N=2 BI action via dimensional reduction. Indeed, the 
six-dimensional action depends upon  its gauge fields via their field strength
only, while one can identify $A_4+iA_5=a$. Hence, the dimensionally reduced
action actually depends upon the derivatives of $a$, and not upon $a$ itself, 
though it is not manifest in eq.~(10). Similarly, the spinor components 
$\j^i_{\a}$ of $W$ in eq.~(8) are supposed to be the Goldstone fermions 
associated with two spontaneously broken (non-linearly realized) 
supersymmetries in four dimensions, $\d\j^i_{\a}=\l^i_{\a}+\ldots$, 
where $\l^i_{\a}$ are the (rigid) spinor parameters.
  
Spontaneously broken symmetries determine the associated Goldstone 
action. Since, in our case, the N=2 BI action is fixed by
the non-linear constraint (13), there should be a 
relation between the
non-linear transformations in question and the constraint (8).
It is now not difficult to find the relevant (without spacetime derivatives) terms 
in the N=2 superfield transformation laws,
$$ \d X=2\L W~,\quad \d W=\L\left(1-\fracm{1}{4}\bar{D}^4\bar{X}\right)
-\fracmm{X}{W}\bar{D}^4\left(\bar{W}\bar{\L}\right)+\ldots~,\eqno(15)$$
where $\L$ is the spacetime-independent (rigid) N=2 superfield parameter,
$$ \L=\l + \q^{\a}_i\l^i_{\a}+\fracm{i}{8}\q^{\a}_i(\s^{\m\n})\du{\a}{\b}
\q^i_{\b}\l_{\m\n}~~,\eqno(16)$$
$X$ is the iterative (to all orders in $W$ and $\bar{W}$, but modulo ambiguous 
$\pa W$- and $\pa\bar{W}$-dependent terms)
 solution to the non-linear constraint (13). The dots in eq.~(15) stand for
the $\pa W$-dependent 
 terms needed for the consistency with the second equation (7). Since those 
terms are ambiguous in the N=2 BI action, I ignore them both in the action and 
 in the transformation laws for simplicity. The invariance of 
the action (12) under the transformations (15) follows from the fact that
$$ D^4W~,\quad (D^3)^{\a}_iW\quad {\rm and}\quad D_{\a\b}W \eqno(17)$$
are all total derivatives in $x$-space, because of eqs.~(8) and (9). The second
relation (15) now follows from the first one by varying the constraint (13). 
Comparing eqs.~(8) and (15) shows that $\l$ is the rigid parameter of broken
translations, whereas $\l^i_{\a}$ are the rigid parameters of two broken
supersymmetries.  Surprisingly enough, there exists yet another non-linear
symmetry with rigid (real and antisymmetric tensor)
parameter
 $\l_{\m\n}$, which is apparently related to
the Goldstone nature of the Maxwell field itself.

It is possible to rewrite the action (12) into the `free-field' form
(subject to the non-linear constraint)
$$ S = \fracm{1}{2}\int d^4x d^4\q\,W^2 + \fracm{1}{8}\int d^4x d^8\q\,
\bar{X}X~,\eqno(18)$$
where I have merely substituted the constraint (13) into eq.~(12) and used
eq.~(6). Equation (18) is the N=2 analogue to the known `free-field' form of 
the N=1 BI action, given by  the 
sum of free actions for an 
N=1 vector multiplet
and  an 
 N=1 chiral multiplet, related by a non-linear constraint \cite{n1,tr}. 
There is, however,  an 
obvious difference between the N=1 and N=2 `free-field'
actions, because the second term in eq.~(18) gives rise to higher derivatives
in components. The existence of the field redefinition eliminating the higher
derivatives is guaranteed by the existence of the  equivalent (gauge-fixed)
D3-brane action without higher derivatives but with non-manifest 
(non-linearly realized or `deformed') unbroken N=2  supersymmetry \cite{my,tr}.

\section{Outlook}

The N=2 BI theory  can be made (rigidly) N=2 superconformally invariant by 
modifying the constraint (13) as ({\it cf.} ref.~\cite{kuz})
$$ X = \fracmm{1}{4}\fracmm{X}{\F^2}\bar{D}^4\left(\fracmm{\bar{X}}{\bar{\F}^2}
\right)+W^2~,\eqno(19)$$
where the N=2 `superconformal compensator' $\F$ is an N=2 restricted chiral
superfield obeying the constraints (7). The original N=2 BI action is obtained
from eqs.~(12) and  (19) by `freezing' $\F$ to $\F=1/\sqrt{b}=(2\p\a')^{-1/2}$.

The bosonic BI lagrangian (sect.~2) interpolates between the Maxwell 
lagrangian, $-\fracm{1}{4}F^{\m\n}F_{\m\n}\equiv -\fracm{1}{4}F^2$, for small 
$F$ and the total derivative, 
$\fracm{i}{8}\ve^{\m\n\l\r}F_{\m\n}F_{\l\r}\equiv \fracm{i}{4}F\tilde{F}$, 
for large $F$ \cite{tr}. In the $b\to 0$ limit the BI lagrangian reduces to
$$ \fracmm{F^2}{F\tilde{F}}~~,\eqno(20)$$
whose N=2 supersymmetric extension \cite{fl}
$$ \int d^8\q\, \fracmm{W^2\bar{W}^2}{K\bar{K}}\left(\fracmm{K+\bar{K}}{K-
\bar{K}}\right) \eqno(21)$$
follows from eq.~(11) in the $b\to 0 $ limit.

One may, therefore, think of $\F$ as a {\it constant} non-covariant background
containing a constant antisymmetric tensor $B_{\m\n}$ on the place of 
$F_{\m\n}$  in eq.~(8). The $b\to 0$ limit is then described by sending $\F$ 
to infinity, at large $B_{\m\n}$ in particular. The N=2 BI action in this
limit is believed to be equivalent to a rank-one (Maxwell) {\it noncommutative}
N=2 supersymmetric gauge field theory via Seiberg-Witten map \cite{sw}, with
$B_{\m\n}$ being the measure of noncommutativity in $x$-space,
$$ \[ x^{\m},x^{\n} \] = i(B^{-1})^{\m\n}~.\eqno(22)$$

\section*{Acknowledgement} It is my pleasure to thank Professor S. J. Gates Jr.
  for kind hospitality extended to me at the University of Maryland in College
Park during preparation of this paper, and Professor E. Ivanov for discussions.

\end{document}